\documentclass[12pt]{article}
\usepackage{epsfig,enumerate,verbatim}
\usepackage{graphicx}
\begin{document}

\begin{center}
{\large \bf Nonlinear Quantum Cosmology}
\end{center}
\vspace{0.1in}

\begin{center}

{Le-Huy Nguyen and Rajesh R. Parwani\footnote{Email: parwani@nus.edu.sg}}

\vspace{0.3in}

{Department of Physics,\\}
{National University of Singapore,\\}
{Kent Ridge,\\}
{ Singapore.}

\vspace{0.3in}

\end{center}
\vspace{0.1in}
\begin{abstract}
We study the effects of an information-theoretically motivated nonlinear correction to the Wheeler-deWitt equation in the minisuperspace scheme for flat, $k=0$, Friedmann-Robertson-Walker (FRW) universes. When the only matter is a cosmological constant, the nonlinearity can provide a barrier that screens the original Big Bang, leading to the quantum creation of a universe through tunneling just as in the $k=1$ case. When the matter is instead a free massless scalar field, the nonlinearity can again prevent a contracting classical universe from reaching zero size by creating a bounce. Our studies here are self-consistent to leading order in perturbation theory for the nonlinear effects.
\end{abstract}

\vspace{0.5in}

\section{Introduction}
When extrapolated back to early times, classical cosmologies typically lead to singularities and a hence breakdown
of predictability in our physical laws. Consider the Einstein-Hilbert action for a  FRW universe,
\begin{equation}
S=\int dt L = {1 \over 2} \int dt N a^3 \left[ {-\dot{a}^2 \over N^2 a^2} + 
{\dot{\phi}^2 \over N^2} -V(\phi) + {k \over a^2} \right]
\end{equation}
where $a(t)$ is the dimensionless\footnote{The physical scale factor $a_{phys}=\beta a$, where $\beta = \sqrt{4 \pi l_{p}^2 \over 3V}$, $l_p$ being the Planck length and $V$ is the volume of the spatial hypersurface divided by $a^3$ \cite{Pinto}.} scale factor, $V(\phi)$ the potential energy of the scalar field $\phi(t)$, $N$ the lapse function, $k=0,\pm 1$, and we have taken $\hbar=c=1$. For slowly varying scalar fields one may ignore the kinetic energy term and set $1/V = a_{0}^2$, so
\begin{equation}
S=\int dt L = {1 \over 2} \int dt N \left[ {-\dot{a}^2 a\over N^2 } + a(k- {a^2 \over a_{0}^2} ) \right] \label{frw1} \, .
\end{equation}
Varying with respect to $N$ and choosing the $N=1$ gauge gives us the Friedmann equation for an isotropic and homogeneous universe with the cosmological constant $\Lambda =3/a_{0}^2$ modeling inflationary sources in the early universe,
 \begin{equation}
 \dot{a}^2 + (k-a^2/a_{0}^2) =0 \, . \label{fried}
\end{equation}
The flat, $k=0$, geometry has the expanding classical solution
\begin{equation}
a=\exp(t/a_0) \label{k0}
\end{equation}
which implies an arbitrarily small universe, $a \to 0$, at early times $ t \to -\infty$. 

Of course when the universe is small there is no reason to expect classical physics to be accurate. At such scales quantum effects should be important and one hopes that they cure classical singularities if and when they occur. In quantum cosmology studies \cite{ Halliwell}, the Wheeler-deWitt (WDW) equation is a functional Schrodinger equation that provides one with a wavefunctional to compute quantum effects in the universe. In the minisuperspace scheme the quantisation is applied to restricted situations such as the FRW model leading to an ordinary Schrodinger equation: From (\ref{frw1}) one gets the canonical momentum $p={\partial L \over \partial \dot{a}}
=-\dot{a}a$. Promoting this to an operator, $p \to \hat{p} = -i {\partial \over \partial a}$
and applying the quantisation to the Friedmann equation (\ref{fried}) leads one to the WDW equation in minisuperspace \cite{Atkatz},
\begin{equation}
\left[ -{\partial^2 \over \partial a^2} + V(a) \right] \psi(a) =0 \, , \label{wdw1}
\end{equation}
which is just a one-dimesional, $0 \le a < \infty$, time-independent Schrodinger equation 
for a particle of mass $m=1/2$ moving in a potential $V(a)=a^2(k-a^2/a_{0}^2)$, with units $\hbar=1$. 

For a spherical geometry, $k=1$, the potential barrier between  $a=0$ and $a=a_0$ means that there is quantum tunneling \cite{Vilenkin,Atkatz}, allowing a classical universe to emerge and exist at finite size ($a \ge a_0$).  However in this case the zero size, $a=0$, universe is actually already avoided in the classical dynamics, Eq.(\ref{fried}), as the classical potential creates a bounce for a classical universe starting from $a>a_0$ and evolving backwards. 

The classical dynamics is different when $k=0$, as seen in (\ref{k0}), because there is then no barrier to prevent a backward evolving universe from collapsing to zero size. Standard quantisation of the $k=0$ geometry does not help: When one chooses  
the outgoing wavefunction as a solution to (\ref{wdw1}), representing an expanding universe, one still encounters the $a=0$ possibility in the sense that the wavefunction allows a universe of arbitrarily small size to exist, see Section(2).

It is generally believed that the functional WDW equation is only an approximate, perhaps semi-classical, description of quantum gravitational effects, partly because of ambiguities in the functional fomalism. For those and other reasons, various modified WDW equations have been investigated, obtained, for example, through a postulated non-commutativity \cite{NC} or by using ideas inspired by loop quantum gravity \cite{Bojo, Ash}. In the latter, a discretised WDW equation emerges which is found to avoid  classical singularities through a bounce.

It is as yet unknown which, if any, of the suggested modifications to classical general relativity is an accurate description of likely new physics at the Planck scale. We therefore adopt in this paper the ``maximum uncertainty (entropy) principle" \cite{Jaynes}, or information-theoretic perspective, to modify the WDW equation and study the consequences. 
The philosophy of the information-theoretic approach is that one should minimise any bias when choosing probability distributions, while still satisfying relevant constraints \cite{Jaynes}.

The maximum uncertainty approach is well known in statistical mechanics \cite{Jaynes}.
For example, if one is only provided with the mean energy  $E = \int \epsilon (x) \ p(x) \ dx$ of a classical statistical system, but not the probability distribution $p(x)$, then one may infer the probability distribution by maximising the Gibbs-Shannon entropy $I_{GS} = -\int p(x) \ln p(x) \ dx$ under the contraint of the given mean energy. That is, one varies $I_{GS} - \beta E$ with respect to $p(x)$ with $\beta$ the lagrange multiplier. This gives the canonical probability distribution $p(x) \propto \exp(-\beta \epsilon (x))$. Of course one must motivate the entropy (uncertainty) functional that is adopted, for example through an axiomatic construction as was done by Shannon \cite{Jaynes}.  

The above approach is now used in many fields and it may also be used to motivate the usual Schrodinger equation \cite{Sch,Sch2} as a one parameter extension of classical ensemble dynamics, arising through a simultaneous minimisation of the Fisher information measure. The Fisher measure,
\begin{eqnarray}
I_F &=& \int {1 \over p} \left( {\partial p \over \partial x} \right)^2 \ dx \ dt \, \label{fish}
\end{eqnarray}
is the simplest information (or inverse uncertainty) measure which satisfies axioms suitable for the ensemble dynamics context \cite{Sch2}, just as the Gibbs-Shannon measure is the simplest measure suitable for classical statistical mechanics. Deformations of the Fisher measure will then give rise to generalisations of the Schrodinger equation suitable for probing potential new physics at shorter distances. It turns out that such deformations lead naturally to nonlinear extensions of Schrodinger's equation \cite{RP1}.  The information theory approach also allows for systematic improvement of the models as our knowledge progresses.

As we saw above, in the minisuperspace scheme Eq.(\ref{wdw1}) is just the time-independent Schrodinger equation defined on the half-line $0 \le a \le \infty$. We postulate therefore that the unknown new physics at shorter distances may be modeled, within the information theory framework, by a nonlinear correction as found previously in Ref.\cite{RP1}, so that the modified equation becomes 
\begin{equation}
\left[ -{\partial^2 \over \partial a^2} + V(a) + F(p) \right] \psi(a) =0 \label{nl1}
\end{equation}
where
\begin{eqnarray} 
  F(p) &\equiv& Q_{1NL} - Q \, , \label{F} 
\end{eqnarray}
with 
\begin{equation}
Q_{1NL}= {  1  \over 2 L^2 \eta^4}  \left[ \ln {p \over (1-\eta) p + \eta p_{+} } + 1 - {(1-\eta) p \over (1-\eta) p + \eta p_{+}} - {\eta p_{-} \over (1-\eta) p_{-} + \eta p} \right]  \,  \label{Q2}
\end{equation}
and 
\begin{eqnarray}
Q &=&  -  {1 \over \sqrt{p}} {\partial^2 \sqrt{p} \over \partial a^2} \, \; . \label{pot1} 
\end{eqnarray}
Here $p(a) = \psi^{\star}(a) \psi(a)$ and $p_{\pm}(a)  \equiv  p(a \pm \eta L)$.
 As we have made $a$ dimensionless, $L>0$ is the dimensionless (scaled by $\beta$) nonlinearity scale and $0 < \eta < 1$ is a parameter that labels a family of nonlinearisations\footnote{As the equation (\ref{nl1}) is only defined on the half-line, we do not symmetrise the nonlinearity with respect to $L$ \cite{RP1}, using the the simplest version here.} .

At the level of the action, $F(p)$ is obtained by varying the Kullback-Liebler (KL) information measure $I_{KL}$ which in the limit $L \to 0$ reduces to the Fisher information measure responsible for the usual linear Schrodinger equation \cite{RP1}. Here is a brief review of the basic KL measure given by 
\begin{equation}
I_{KL}(p,r) = -\int p(x) \ln {p(x) \over r(x)} \ dx \, . \label{kl}
\end{equation}
It measures the relative uncertainty between two probability distributions, $p(x)$ and a reference $r(x)$. If we have no useful {\it a priori} information then $r(x)$ can be taken to be a uniform distribution and the KL measure then reduces to the Gibbs-Shannon entropy. On the other hand 
if in Eq.(\ref{kl}) one chooses the reference  distribution $r(x)$ to be the same as $p(x)$ but with infinitesimally shifted arguments, that is $r(x) = p(x + \Delta  x)$, then to lowest order,
\begin{eqnarray}
I_{KL} ( p(x), p(x + \Delta(x)) &=& {- (\Delta x)^2  \over 2} I_F (p(x)) + O(\Delta x)^3  \, . \label{connection}
\end{eqnarray}
So to lowest order, minimising the Fisher information, which leads to the linear Schrodinger equation,  is the same as maximising the KL uncertainty measure for two probability distributions that are  close to each other, $r(x) = p(x + \Delta x)$; one might interpret $\Delta x$ as the resolution at which the coordinates become distinguishable \cite{RP1}. Thus the KL measure interpolates between the simplest inference measures used for quantum theory and statistical mechanics.

Furthermore, the (regularised) Kullback-Liebler measure used to obtain (\ref{nl1}) is arguably the simplest generalised measure which has properties similar to those of the Fisher measure and yet leads to nonsingular equations of motion.  In particular, the nonlinear correction $F$ preserves the invariance of the Schrodinger equation to changes in the norm of the wavefunction, $\psi \to \lambda \psi$. We refer the interested reader to \cite{Sch,RP1} for further details.

In summary, the motivation for choosing the regularised KL measure, and hence the family of nonlinear equations (\ref{nl1}), is that it leads to a minimal deformation of the Schrodinger equation within the information theory framework. As stated above, we intend to study the consequences of these nonlinear WDW equations, focusing in this paper on some simple situations.  In the next section we will show that the nonlinearly corrected quantum dynamics can avoid the zero size, $a=0$, possibility that is present in an evolving flat FRW universe with a cosmological constant, Eq.(\ref{k0}). Then in Section(3) we study a flat FRW universe where the matter is represented by a free massless scalar field. Here we show that the original classical singularity can be avoided by the modified  classical equations induced by the nonlinear correction to the quantum dynamics.  

We remark that the issue of singularity avoidance in cosmology has been previously studied by several authors using different approaches, see for example the reviews \cite{Bojo,Ash,Sing} and references therein. What we are proposing here is a different framework, that of information theory, to motivate any corrections to the dynamics of the universe. As discussed in the concluding section, we hope to use this framework to also study other issues in cosmology and to see if this method of ``minimum bias" can be related to other approaches in the literature.

\section{Cosmological Constant}
The flat, $k=0$, universe is favoured by current data so it is of interest to see if the nonlinear correction helps screen the initial singularity typical for such FRW universes. Here we study the case when matter is only in the form of a cosmological constant. 
Setting $k=0$ and $a=lb$, with $l=a_{0}^{1/3}$, in (\ref{nl1}) we get 
\begin{equation}
\left[ -{\partial^2 \over \partial b^2} -b^4 +l^2  F(p(lb)) \right] \phi(b) =0 \, ,
\end{equation}
$\psi(a) \equiv \phi(b)$. We assume that the nonlinearity is small even in the early universe so that $F$ may be expanded perturbatively in $L$ to lowest order,
\begin{equation}
F(b) = {\eta(3-4\eta) L \over l^3 } f(b) + O(L^2) \, ,
\end{equation} 
where
\begin{equation}
f(b)={q' \over 12 q^3} (2 q'^2 -3q''q) \,  \label{f1est}
\end{equation}
and $q(b) =\phi^*(b)\phi(b)$. Thus we may write 
\begin{equation}
\left[ -{\partial^2 \over \partial b^2} -b^4 + \eta(3-4\eta) \epsilon f(b) \right] \phi(b) =0 
\end{equation}
where $\epsilon \equiv L/l$ measures the strength of the nonlinearity. Assuming 
$\epsilon \ll 1$, we may solve the equation by iterating about the unperturbed solution for $\epsilon =0$. The unperturbed solution which represents an expanding universe at large scale is given by a Hankel function
\begin{eqnarray}
\phi_0(b) &\propto&  \sqrt{b} H_{1/6}^{(2)} (b^3/3) \\
&\sim& \sqrt{{6 \over \pi b^3}} \exp{\left[ -i(b^3 -\pi)/3\right]} \,\,\,  \mbox{as} \; \; b \to \infty \, .
\end{eqnarray}
Recalling $p=-a \dot{a}$, the negative momentum of this solution corresponds to $\dot {a} >0$ as required.
From this unpertubed solution one calculates $q_0=\phi_{0}^{*} \phi$ and then $f_0(b)$ which is shown in Fig.(1). Then to leading nontrivial order in the iteration one has a linear Schrodinger equation with an effective potential 
\begin{equation}
V_{eff} = -b^4 + \eta(3-4\eta) \epsilon f_0(b)  \, .
\end{equation}
For small $b$, 
\begin{equation}
f_0 \approx 0.1 b \label{f0}
\end{equation}
and so for $\eta <3/4$ there is an effective potential barrier, a finite size universe coming into being through quantum tunneling. 

It is interesting to calculate the probability for tunneling through the barrier. For the $k=1$ case the result in the WKB approximation is known in the linear theory \cite{Atkatz},
\begin{equation}
P_{k=1} = \exp \left[-2 a_{0}^2 /3\right] \, ,
\end{equation}
with small nonlinear corrections not affecting the result significantly, so the universe is likely to be born with a small size \cite{Atkatz}. 
For the present $k=0$ case the WKB formula applied to the effective Schrodinger equation gives
\begin{equation}
P_{k=0} \approx \exp\left[-2 \int_{0}^{b_1} db \sqrt{V_{eff}(b)} \right]
\end{equation}
with $b_1$ the point where $V_{eff}(b_1)=0$. Using the expression (\ref{f0}) we may estimate 
\begin{equation}
b_1 = 0.46(\eta (3-4\eta)\epsilon)^{1/3}
\end{equation}
and hence 
\begin{equation}
P_{k=0} \approx \exp{\left[-0.1 \eta (3-4\eta) \epsilon) \right]} \, ,
\end{equation}
so that for fixed $\eta < 3/4$ one may say that small values of $\epsilon$ are ``preferred", a conclusion which is self-consistent with our approximation  $\epsilon \ll 1$\footnote{The $k=-1$ results will be discussed elsewhere.}.

Having obtained the effective potential above, one may alternatively discuss the modified classical dynamics of the universe. It is clear now that for parameter values $\eta < 3/4$, a backward evolving classical universe will experience a bounce instead of shrinking to zero size.

\section{Free Massless Scalar Field}
As the WDW equation is independent of time, so in quantum cosmology one often looks at correlations between variables to describe the evolution of the universe \cite{Halliwell}. One possibility is to use a free massless scalar field as an internal clock \cite{Ash}, which we also do below. The coupling of a free massless scalar field to gravity may also be viewed as the other extreme to the model we studied in the last section; here we have kinetic energy but no potential energy for the matter.

Following convention\footnote{Actually we need this transformation to get the Klien-Gordon equation (\ref{wdw2}) with conventional kinetic terms for both variables so that we may implement the nonlinearisation as in Ref.\cite{RP1}.},  we define $\alpha = \ln a$ so that the $k=0$ classical FRW action becomes
\begin{equation}
S=\int dt L = {1 \over 2} \int dt \ N e^{3 \alpha} \left[ {-\dot{\alpha}^2 \over N^2 } + 
{\dot{\phi}^2 \over N^2}  \right] \, ,  \label{act2}
\end{equation}
giving the following classical equations on varying with respect to $\alpha, \phi, N$ and then setting $N=1$,
\begin{eqnarray}
\ddot{\phi} + 3 \dot{\phi} \dot{\alpha} &=& 0 \, , \\
2 \ddot{\alpha} + 3 \dot{\alpha}^2 + 3\dot{\phi}^2 &=& 0 \, , \\
-\dot{\alpha}^2 + \dot{\phi}^2 &=& 0 \, , \label{const1}
\end{eqnarray}
the last equation being a constraint. After a scaling of the time variable, we have a solution 
\begin{eqnarray}
\alpha &=& { 1 \over 3} \log t + C' \, ,  \label{class1} \\
\phi &=& {1 \over 3} \log t  \label{class2}
\end{eqnarray}
where $C'$ is an arbitrary constant (we have only indicated the classical solution which interests us, for which $\alpha$ increases as the internal time $\phi$ increases). As $t \to 0$, $\alpha \to -\infty$ and so $a \to 0$, the classical universe arises from an initial singularity.

To quantise the classical theory, we start as usual from the Hamiltonian corresponding to (\ref{act2}),
\begin{equation}
H= {N e^{-3 \alpha} \over 2} \left[-p_{\alpha}^{2} + p_{\phi}^{2} \right] 
\end{equation}
where the canoical momenta are $p_{\alpha} = {-e^{3 \alpha} \dot{\alpha} \over N}$ and $p_{\phi} = { e^{3 \alpha} \dot{\phi} \over N}$. The classical constraint (\ref{const1}) becomes the Hamiltonian constraint $\hat{H} \psi =0$ giving us the WDW equation 
\begin{equation}
\left[-{\partial^2 \over \partial \phi^2}+{\partial^2 \over \partial \alpha^2} \right] \psi(\phi, \alpha) =0 \, . \label{wdw2}
\end{equation}
The general solution of (\ref{wdw2}) can easily be obtained by separation of variables and then a wavepacket constructed \cite{Kiefer,Pinto},
\begin{equation}
\psi = \int_{-\infty}^{\infty} w(k) \ A_k(\alpha) B_{k}(\phi)
\end{equation}
where $w(k)$ is an arbitrary function of $k$ and 
\begin{eqnarray}
A_k ({\alpha}) &=& a_1 \exp({ik\alpha}) + a_2 \exp({-ik\alpha}) \, , \\
B_k ({\phi}) &=& b_1 \exp({ik\phi}) + b_2 \exp({-ik\phi}) \,  , 
\end{eqnarray}
for some constants $a_1,a_2,b_1,b_2$. We would like the state to represent a large universe when the intrinsic time $\phi$ is large, and so we construct such a localised wavepacket by taking $a_2=b_1=0$ and the Gaussian weight $w(k) = \exp \left[ -{(k-d)^2 \over \sigma^2} \right]$ to get 
\begin{equation}
\psi = a_1 b_2 \sigma \sqrt{\pi} \exp \left[ -{(\alpha-\phi)^2 \over 4} \right] \exp(id(\alpha-\phi)) \, ,
\end{equation}
$\sigma$ and $d$ being constants. The density 
\begin{equation}
p(\phi, \alpha) \equiv \psi^{*} \psi = (a_1 b_2 \sigma)^2 \pi \exp \left[ -{(\alpha-\phi)^2 \over 2} \right] \label{un2}
\end{equation}
is clearly localised near $\alpha \approx \phi$ for large $\phi$; we take this to represent a classical universe at large time. But we also see that the universe would have been arbitrarily small, $\alpha \to -\infty$, in the distant past $\phi \to -\infty$. 

We now proceed to nonlinearise the Klien-Gordon equation (\ref{wdw2}) following the information-theoretic approach discussed in Ref.\cite{RP1} to get 
\begin{equation}
\left[-{\partial^2 \over \partial \phi^2}+{\partial^2 \over \partial \alpha^2} -F_{\alpha}(p) + F_{\phi}(p) \right] \psi(\phi,\alpha) =0 \, . \label{nl2}
\end{equation}
where $F_{\alpha}$ and $F_{\phi}$ have the same form as the $F$ in (\ref{F}) but with generally distinct nonlinear parameters $L_{\alpha}>0$ and $L_{\phi}>0$ corresponding to the gravitational and matter degrees of freedom.

As in the previous section, we solve (\ref{nl2}) by perturbation and iteration to lowest nontrivial order using (\ref{un2}) to get an effective linear equation 
\begin{equation}
\left[-{\partial^2 \over \partial \phi^2}+{\partial^2 \over \partial \alpha^2}  + V_{eff} \right]  \psi(\phi, \alpha) =0 \,  \label{eff2}
\end{equation}
with 
\begin{eqnarray}
V_{eff} &=& u \left[ \sigma^2 (\phi -\alpha)^2 -3 \right] (\phi-\alpha) \, , \label{veff} \\
u &\equiv& { \eta (3-4\eta) (L_{\alpha} + L_{\phi} ) \sigma^4 \over 12} \, .
\end{eqnarray}
This linear equation describes the approximate, to leading nontrivial order in perturbation theory, quantum dynamics of the states that are still highly localised.

As the hyperbolic nature of the Klien-Gordon equation makes the tunneling analogy problematic, see for example Ref.\cite{multi}, we proceed differently. We obtain the effective classical equations that imply, through the correspondence principle, the modified quantum equation (\ref{eff2}). First we note that the modified equation (\ref{eff2}) corresponds to the Hamiltonian 
\begin{equation}
\hat{H} = {N e^{-3\alpha} \over 2 } \left[-\hat{p}_{\alpha}^{2} + \hat{p}_{\phi}^{2} +V_{eff}(\phi,\alpha) \right] 
\end{equation}
and this arises from canonically quantising the classical action 
\begin{equation}
S=\int dt L = {1 \over 2} \int dt N e^{3 \alpha} \left[ {-\dot{\alpha}^2 \over N^2 } + 
{\dot{\phi}^2 \over N^2} - e^{-6\alpha} V_{eff}(\phi) \right] \, .
\end{equation}
This effective classical action gives the following modified evolution equations (in the $N=1$ gauge)
\begin{eqnarray}
\ddot{\phi} + 3 \dot{\phi} \dot{\alpha} + {1 \over 2 } e^{-6\alpha} {\partial V_{eff} \over \partial \phi} &=& 0 \, , \label{m1} \\
2 \ddot{\alpha} + 3 \dot{\alpha}^2 + 3\dot{\phi}^2 + e^{-6\alpha} \left[ 3 V_{eff} - {\partial V_{eff} \over \partial \alpha} \right] &=& 0 \, , \label{m2} \\
-\dot{\alpha}^2 + \dot{\phi}^2 +  e^{-6\alpha} V_{eff} &=& 0 \, . \label{const2}
\end{eqnarray}
Note that 
\begin{equation}
{\partial V_{eff} \over \partial \phi} = -{\partial V_{eff} \over \partial \alpha} =3u \left[ \sigma^2 (\phi -\alpha)^2 - 1 \right] \, . \label{note1}
\end{equation}

It is easy to verify that the modified constraint equation (\ref{const2}) combined with any of the other two evolution equations (\ref{m1},\ref{m2}) implies the third.  Without loss of generality in the subsequent analysis we set $\sigma=1$. 

We remark that since the quantum states used to obtain the effective potential (\ref{veff}) were localised, the effective classical equations  (\ref{m1}-\ref{const2}) are a self-consistent description of the mean dynamics of such states.

\subsection{Analytical Results}
A subset of solutions to the new coupled equations (\ref{m1}-\ref{const2}) can be studied analytically by assuming the correlation $\alpha =\phi$ at all times. The constraint equation is then automatically satisfied and the other equations reduce to 
\begin{equation}
3 \dot{\alpha}^2 + \ddot{\alpha} -{3 u \over 2} e^{-6 \alpha} =0
\end{equation}
which is easily integrated to give 
\begin{equation}
e^{6 \alpha} \ \dot{\alpha}^2 = 3 u \alpha + C \label{analy}
\end{equation}
where $C$ is the integration constant. Since the left-hand-side is positive, therefore taking $\eta < 3/4$ as before (that is, $u>0$) gives
\begin{equation}
\alpha \ge {- C \over 3 u}  \label{amin}
\end{equation}
and so 
\begin{equation}
a \ge a_{min} = \exp({-C \over 3u}) 
\end{equation}

That is, there exists a class of solutions for which $\alpha(t) \equiv \phi(t)$ for all $t$: Such universes have a minimum nonzero size, which depends on the initial conditions that fix $C$, and hence the Big Bang singularity is avoided, being replaced by a bounce. Note that while a $\alpha \equiv \phi$ correlation is also possible  in the original classical dynamics, (\ref{class1},\ref{class2}), only in the modified dynamics is the singularity avoided. 

The $\alpha \equiv \phi$ solutions of the modified dynamics are stable to small perturbations as one can see by setting $\alpha(t) = \phi + \delta(t)$ and linearising the dynamical equations for $\delta \ll 1$ to get
\begin{equation}
\ddot{\delta} + {9 u \delta \over 2 a^6(t) } =0
\end{equation}
which is the equation for an oscillator with time-dependent frequency; $\delta$ is thus bounded.

It should be emphasised that though the $\alpha \approx \phi$ solutions to  (\ref{m1}-\ref{const2}) form a subset of solutions to those equations, they are a very important class because the effective potential (\ref{veff}) was constructed precisely from quantum states (\ref{un2}) which were localised around $\alpha \approx \phi$. Furthermore, to reiterate, while in the original dynamics such states encounter a singularity, in the modified dynamics one sees instead a bounce.

\subsection{Numerical Solution I}
We now investigate the coupled effective classical equations (\ref{m1}-\ref{const2}) numerically.  We show in the Appendix that if one solves the two second-order equations (\ref{m1}-\ref{m2}) with initial conditions at $t=t_0$ that satisfy the constraint equation (\ref{const2}), then the constraint equation is satisfied for all times.  The initial configuration is chosen to be a large universe ($a > 1$) at large time $t$. Notice that the nonlinear corrections come with the pre-factor $\exp(-6\alpha) $ which causes the nonlinear terms to be negligible when the universe is large. So our initial configuration satisfies very closely the classical, unmodified, FRW equations which we had considered earlier (\ref{class1},\ref{class2}). 

Consider first the choices, of initial time $t_0 =27$ and $C'=0$. This gives  initial conditions $\alpha_0 =\phi_0 =\log 3$ and $\dot{\alpha}_0 = 1/81$, with $\dot{\phi}(t_0)$ determined by the constraint equation. We set $u=0.01$. The numerically solved equations give the plot shown in Fig.(2). The solution is found to satisfy $\alpha =\phi$ for all time with $\alpha_{min} \approx -2.5$. In the figure we also compare the modified trajectory with the classical result, showing a bounce in the first case and a singularity in the latter.

In the Appendix we compare this numerical solution with the analytical results from the previous subsection.

\subsection{Numerical Study II}
 By choosing nonzero values of $C'$ in (\ref{class1}) we can study more general initial conditions for which $\alpha \neq \phi$. We keep $t_0=27$ and $u=0.01$. Figs.(3-6) show some examples, in all cases a bounce occurs at some nonzero size that depends on the initial conditions.
 
As we keep increasing $\alpha_0$ beyond $\log(3.5)$, while keeping $t_0=27$ fixed, the curve for $\alpha(t)$ shows an increasingly  sharper turning point and increasingly lower minimum. This might seem to indicate that a 
singularity, $\alpha \to -\infty$, will  eventually appear when starting from some initial conditions corresponding to large $\alpha_0$ for a fixed $t_0$. However note that in those cases where  $\alpha$ becomes increasingly negative, the corrections from $V_{eff}$ to the classical equations  have become large, recall the prefactor $\exp(-6 \alpha)$, and we are thus beyond the validity of our perturbative approximation. Therefore all that one can conclude at this point is that within the regime where the perturbative approximation is valid, the singularities are resolved, being replaced by a bounce.

\section{Conclusion}
We have used an information-theoretic approach to model the unknown new physics that is generally believed to exist near the Planck scale. For simplicity, the resulting nonlinear WDW equation in minisuperspace was analysed assuming the nonlinearity to be weak. The nonlinearity in the WDW equation induces an effective potential in the modified classical dynamics: In the modified $k=0$ FRW models with a cosmological constant or a free massless scalar field, a classically contracting universe may experience  a bounce instead of reaching zero size. 

The nonlinearisation of the quantum equation makes the Hamiltonian, and hence the effective classical potential, state dependent. The state-dependence is not necessarily undesirable: It could perhaps be used to select states that lead to mathematically well-defined evolutions and hence hopefully guide us to the effective (emergent) laws that govern reasonable universes like the one we inhabit.

The state-dependence of our bounce is in contrast to some other approaches such as that of loop quantum cosmology \cite{Ash,Bojo}. In passing we note that in loop quantum cosmology the modified WDW equation is a linear {\it difference} equation while in our approach it is a  
nonlinear {\it difference}-differential equation; this seems intriguing.

In the free massless scalar field case, the effective potential that was generated through the nonlinearity was responsible for the bounce and hence also a momentary  acceleration  of the universe. However as we used a perturbative approach, the inflation is mild. It would be of interest to study the nonlinearity non-perturbatively to see if realistic inflation can be generated by the nonlinearity starting from free matter fields, as there are indications that the nonlinearity has novel non-perturbative properties \cite{RP2}. Indeed, one might expect the nonlinearity to be large near the Planck scale and perhaps weaken as the universe expands, so it might be useful to make the nonlinearity parameter $L$ dependent on the scale factor $a(t)$.

As we studied the realisation of the bounce in two extreme cases of matter, pure potential energy and pure kinetic energy, we think that similar conclusions will hold for more general matter though this would require checking. It would be useful to study in a realistic cosmological model not only the singularity avoidance but also the residual long time consequences of the nonlinearity, for example its effect on the cosmic microwave background and its relevance to the present acceleration of the universe. Such phenomenology might help fix, and perhaps also provide physical meaning to, the free parameter $0<\eta<1$ in our model: Recall that only for $\eta <3/4$ did we find a bounce in the perturbative regime. Phenomenology would eventually also be important for discriminating between the predictions of the various different approaches to quantum cosmology in the literature. 

Of course one should go beyond the FRW minisuperspace and study less symmetric geometries that are better models for a beginning universe. Other gravitational systems for which an information-theoretically modified dynamics might yield interesting results are quantum black/worm holes and the current classical accelerating universe.\\

\section*{Acknowledgement}
We thank Sayan Kar for clarifying discussions and the anonymous referees for useful comments.

\section*{Appendix: Numerics}
\subsection*{A Lemma}

Denote the left-hand-side of the constraint (\ref{const2}) by
\begin{equation}
A(t) \equiv -\dot{\alpha}^2 + \dot{\phi}^2 +  V(\alpha,\phi)  
\end{equation}
where we have also defined $V(\alpha,\phi)\equiv e^{-6\alpha} V_{eff}$. Multiplying (\ref{m1}) by $\dot{\phi}$ and (\ref{m2}) by $\dot{\alpha}$ and using those results in $\dot{A}$, one obtains
\begin{equation}
\dot{A} = -3 \dot{\alpha} A
\end{equation}
or
\begin{equation}
e^{3\alpha} A = \mbox{constant}.
\end{equation}
Thus, when solving (\ref{m1},\ref{m2}) simultaneously, if the initial conditions are chosen such that $A=0$ at some time $t_0$, then the constraint equation (\ref{const2}) is satisfied for all time. 

The consistency of the constraint equation may be discussed more generally using constraint algebra arguments, as for example in Ref.\cite{bojo2}.

\subsection*{First Order Constraints}
The numerical solution of (\ref{m1}, \ref{m2}) was obtained using the Maple 11 software. Initially we used the Fehlberg fourth-fifth order Runge-Kutta method and rechecked the results using the forward Euler method.

In addition we performed the following cross-check to detect numerical errors. The two equations (\ref{m1},\ref{m2}) are two second-order equations. By combining them in certain ways we obtain a new first-order constraint as follows: Multiply Eq.(\ref{const2}) by $3/2$, add Eq.(\ref{m1}), subtract Eq.(\ref{m2}) and use  Eq.(\ref{note1}) to get 
\begin{equation}
\ddot{x} + 3 \dot{x}\dot{\alpha} =0
\end{equation} 
where $x=\phi-\alpha$. Integrating the last equation gives
\begin{equation}
e^{3 \alpha} (\dot{\phi}-\dot{\alpha}) = \mbox{constant} .
\end{equation}
Together with the original constraint equation (\ref{const2}), this new first-order constraint can be used for checking the numerical results: The numerical solutions of 
(\ref{m1},\ref{m2}) give us the values of $\alpha, \phi$ and their first derivatives at any point. Such data from several points in our plots were inserted into the two first-order constraint equations to check for consistency. 

\subsection*{Checking a $\alpha=\phi$ solution}
The solution discussed in Section(3.2) was found to have $\alpha=\phi$ for all time. It had $\alpha_{min} \approx -2.5 $ and $t_{min} \approx -1.4$. These values can be compared with the analytical results of Section(3.1). The constant $C$ in ({\ref{analy}) can be evaluated using the data from Section(3.2) to give $C \approx 0.078$ and hence (\ref{amin}) gives $\alpha_{min} \approx -2.6$ in good agreement with the results from the plot Fig.(2). Next, one may integrate (\ref{analy}),
\begin{equation}
\int_{t_0}^{t_{min}}dt = \int_{\alpha_0}^{\alpha_{min}} d \alpha 
{ e^{3 \alpha} \over \sqrt{C + 3 u \alpha} } \, .
\end{equation}
Using the known parameters, this yields $t_{min} \approx -1.4$ again in good agreement with the plot Fig.(2).

\section*{Figure Captions}

\begin{itemize}
\item  Figure 1 : The induced potential $f_0(b)$ in Eq.(19) is always positive; it vanishes at the origin, peaks around $b=1.2$ and approaches zero as the dimensionless scale factor $b$ approaches infinity.

\item  Figure 2 : For the initial conditions described in Section(3.2), $\alpha_0= \log a_0=\log(3)$: The dashed line represents the trajectory of the classical $k=0$ FRW universe with a massless scalar field; it hits the singularity at $t=0$. The solid line is the trajectory in the modified dynamics, showing a bounce. The curve for $\phi$ coincides that for $\alpha$. (Natural units are used as mentioned in Section 1).

\item  Figure 3 : Results for $\alpha_0= \log(2.5)$: The dashed line represents the trajectory of the classical while the solid line corresponds to the modified dynamics.

\item  Figure 4 : Results for $\alpha_0= \log(2.5)$: Trajectories for $\alpha$ and $\phi$ in the modified dynamics. The $\phi$ curve is above the $\alpha$ curve at large positive time but below it for large negative time.

\item  Figure 5 : Results for $\alpha_0= \log(1.5)$: The dashed line represents the trajectory of the classical while the solid line corresponds to the modified dynamics.

\item  Figure 6 : Results for $\alpha_0= \log(1.5)$: Trajectories for $\alpha$ and $\phi$ in the modified dynamics. The $\phi$ curve is above the $\alpha$ curve at large positive time but below it for large negative time.

\end{itemize}

\end{document}